\PassOptionsToPackage{table}{xcolor}
\documentclass[cameraready]{Interspeech}

\title{Geometry-Informed Distributed Acoustic Scene Understanding}

\author[]{Yiyuan}{Yang}
\author[]{Shitong}{Xu}
\author[]{Niki}{Trigoni}
\author[]{Andrew}{Markham}
\address{Department of Computer Science, University of Oxford, United Kingdom}
\email{\{yiyuan.yang, shitong.xu, niki.trigoni, andrew.markham\}@cs.ox.ac.uk}


\usepackage{comment}
\usepackage{amsmath} 
\usepackage{amssymb} 
\usepackage{bm}  
\usepackage{tabularx}

\usepackage{listings}

\lstdefinelanguage{json}{
    basicstyle=\ttfamily\scriptsize,
    numbers=left,
    numberstyle=\tiny,
    stepnumber=1,
    numbersep=5pt,
    showstringspaces=false,
    breaklines=true,
    frame=single,
    backgroundcolor=\color{gray!5},
    stringstyle=\color{blue},
    keywordstyle=\color{red},
    commentstyle=\color{green!50!black}
}

\begin{document}
\maketitle

\begin{abstract}
Acoustic scene understanding in multi-room environments is a difficult task. Most existing systems use a single centralized microphone array, and they often fail because walls and doors block sound signals. To address this challenge, we propose a geometry-informed distributed acoustic scene understanding framework. Our system leverages distributed microphones and uses an audio spectrogram transformer and a topology-aware graph neural network to fuse spatio-temporal acoustic features. Then, these features are decoded into discrete semantic triplets. Finally, a frozen large language model combines these symbolic observations with the environmental geometry. This allows the system to perform spatial understanding, infer plausible missing transitions, and generate a physically consistent narrative of the scene. Experiments on a custom multi-room simulator demonstrate that our framework outperforms centralized baselines and improves spatial consistency under simulated occlusion.
\end{abstract}

\section{Introduction} \label{sec:introduction}

\begin{figure}[!t]
  \centering
  \includegraphics[width=\linewidth]{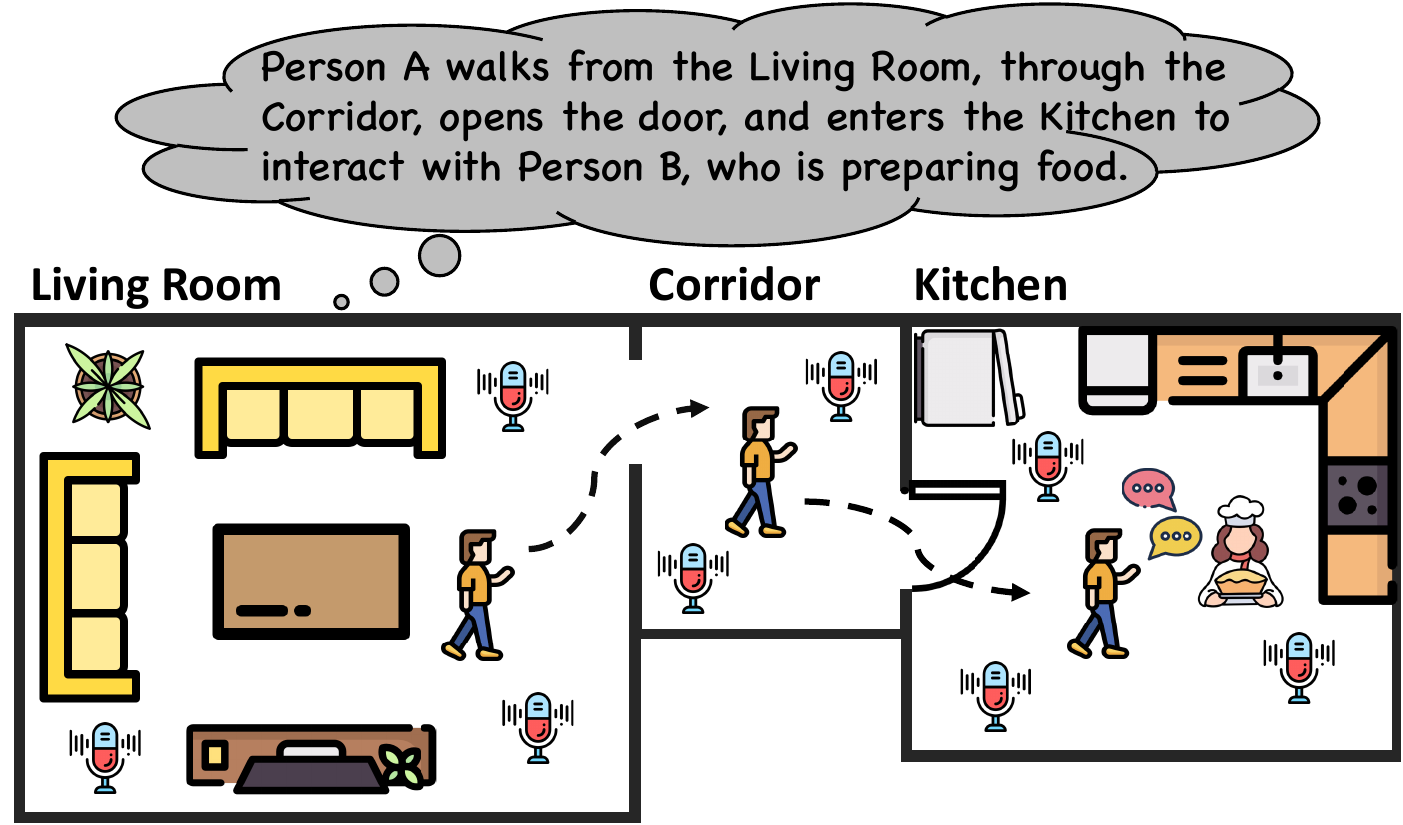}
  \caption{Illustration of the multi-room acoustic scene understanding task. Distributed microphones capture fragmented acoustic observations, which are logically aggregated to generate a coherent scene narrative.}
  \label{fig:task_overview}
\end{figure}

Acoustic scene understanding aims to monitor, identify, and interpret environmental events through auditory signals~\cite{mesaros2016tut}. Compared to vision-based systems, acoustic sensors better protect user privacy and operate effectively in complete darkness or obstructed environments, making them suitable for long-term monitoring and indoor security. For instance, in elderly care, an acoustic scene understanding system can alert family members to a fall or a medical emergency without intrusive constant visual surveillance~\cite{zigel2009method}. However, to be practically viable, such a system must go beyond isolated event detection and instead provide a coherent description of ongoing events~\cite{gong2023listen}. Rather than merely reporting fragmented acoustic tags, the system should construct a continuous narrative that captures the spatiotemporal evolution of a scene, as illustrated in Figure~\ref{fig:task_overview}.

However, realizing this goal in real-world environments is hindered by acoustic occlusion. A primary reason is that current acoustic scene understanding systems often assume a single listening location or acoustic space~\cite{zheng2024bat,jiang2026sci}. In a building with multiple rooms, walls and doors act as physical barriers that block or muffle sound. This creates non-line-of-sight zones where a central acoustic sensor cannot accurately ``hear" events happening in other rooms~\cite{aparicio2024baseline}. For instance, if the system is in the living room, it might miss the resident preparing tea in the kitchen because the wall significantly attenuates the sound. This results in a fragmented and incomplete understanding of the entire scene.

To solve this, researchers have moved toward distributed microphone networks~\cite{rodomagoulakis2017room,vesperini2016neural,yang2025distributed}. By deploying multiple microphones in different rooms, the system can achieve much better coverage. Current methods usually combine the audio data from all acoustic sensors to make a final decision~\cite{ruiz2022distributed,grinstein2023gnn,hou2023relational}. However, these methods often ignore the physical topology of the environment. They treat each detected sound as an isolated data point and fail to consider how sound ``leaks" through a wall or how a person moves between different rooms. Without knowing the geometry, these systems cannot effectively judge which sensor's information is the most reliable when a sound is simultaneously captured across multiple rooms.

Furthermore, another critical challenge in distributed acoustic scene understanding is logical grounding. In a real physical environment, human events must follow spatial rules and constraints. For example, a person cannot move from a bedroom to a kitchen without first passing through a corridor. If the acoustic signal captured by the corridor microphone has a low Signal-to-Noise Ratio and misses footsteps, the system will output non-physical jumps in footstep prediction between rooms. To produce a trustworthy description, the system needs to be aware of the actual geometry so it can infer missing intermediate transitions. This necessitates a reasoning mechanism that utilizes spatial logic to integrate incomplete acoustic observations into a physically plausible reconstruction~\cite{kuan2025can,lee2025audio}.

In this study, we propose a framework for distributed acoustic scene understanding. Our approach integrates physical geometry into the perception process to ensure the narrative is logical and accurate. Specifically, our main contributions are:
\begin{itemize}
    \item We formalize the acoustic scene understanding task in multi-room environments using a structured semantic triplet $\langle \text{subject, relation, object} \rangle$. This extends traditional event detection toward generating logically grounded narratives.

    \item We propose a framework that features a topology-aware graph fusion for distributed acoustic sensing and a geometry-aware Large Language Model (LLM)-based reasoning module. By incorporating geometry constraints, the system can infer missing acoustic information and ensure the narrative adheres to real-world physical rules.

    \item We demonstrate the efficacy of our approach using a multi-room acoustic simulator that models complex indoor propagation. The results show that our framework achieves good accuracy and improves logical consistency in simulated scenes with spatiotemporal discontinuities.
\end{itemize}

\section{Related Work} \label{sec:related_work}

\subsection{Sound Event Localization and Detection}
Sound Event Localization and Detection (SELD) aims to recognize sound categories while simultaneously estimating their spatial positions~\cite{aparicio2024baseline,yang2025efficient}. Recently, Transformer-based architectures such as the Audio Spectrogram Transformer have been introduced to better capture long-range spectral dependencies~\cite{gong21b_interspeech}. At the same time, researchers have explored self-supervised learning frameworks to mitigate the scarcity of labeled spatial acoustic signals. For instance, w2v-SELD pre-trains on unlabeled 3D audio to learn more robust spatial representations~\cite{santos2023w2vseld}. Multi-scale feature fusion modules have also been proposed to jointly model spectral and temporal variations, further improving performance~\cite{mu2024mff}. Furthermore, recent work has started to move from professional microphone arrays toward consumer-level stereo audio~\cite{aparicio2024baseline}. Despite these advances, most SELD frameworks operate within single acoustic spaces and focus on coordinate estimation rather than structured multi-room scene understanding.

\subsection{Distributed and Multi-Room Acoustic Sensing}
Traditional centralized microphone arrays often suffer from signal attenuation in multi-room environments~\cite{grinstein2023gnn}. Physical barriers create non-line-of-sight regions, limiting the reliability of acoustic observations. To address this limitation, distributed microphone networks deploy sensors across multiple rooms~\cite{yang2025distributed}. This setup provides broader spatial coverage and reduces blind spots. At the same time, graph neural networks have been applied to model the physical structure of indoor spaces~\cite{grinstein2023gnn}. In these formulations, microphones are represented as nodes, and edges encode room connectivity or spatial relationships~\cite{hou2023relational}. Graph-based fusion helps the system emphasize direct acoustic propagation paths and suppress spurious noise leakage~\cite{hou2023relational}. Overall, distributed sensing frameworks offer a more complete representation of complex indoor acoustic scenes compared to centralized microphone systems. 

\subsection{Geometry-Informed Audio-Language Reasoning}
Recent research has moved beyond isolated sound event tagging toward constructing structured and interpretable scene understanding. One common approach converts acoustic data into symbolic formats such as semantic tokenizations~\cite{xie2025audioreasoner}. This structured representation provides a foundation for higher-level reasoning. Large audio-language models further integrate audio encoders with pre-trained language models to enable more complex reasoning over acoustic scenes~\cite{kuan2025can,chu2023qwen}. For example, some systems can answer open-ended questions about spatial relationships between sound sources~\cite{zheng2024bat}. In multi-room environments, incorporating geometric information becomes particularly important. Several recent studies have begun exploring geometry-aware reasoning in audio-language settings, although explicit floor-plan-constrained narrative reconstruction remains underexplored~\cite{jiang2026sci}. Some methods also employ tool-augmented reasoning, where external signal analysis modules are invoked to estimate acoustic parameters~\cite{lee2025audio}. By combining perceptual modeling with structured reasoning, these systems move closer to reliable reconstruction of complex indoor events.

\section{Methodology}
\label{sec:methodology}

\begin{figure*}[!t]
  \centering
  \includegraphics[width=1\linewidth]{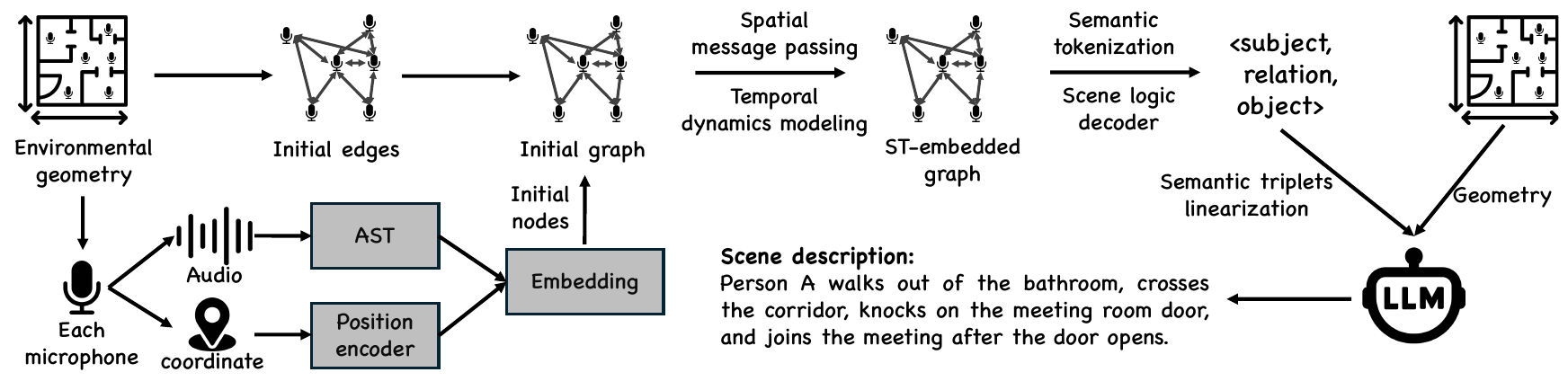}
    \caption{Workflow of the proposed distributed acoustic scene understanding framework. The pipeline converts raw audio signals and environmental geometry into a textual description by first extracting spatio-temporal features across distributed nodes. These features are translated into structured semantic triplets, which serve as a symbolic prompt for a frozen LLM to perform narrative synthesis.}
    \vspace{-10pt}
  \label{fig:workflow}
\end{figure*}

In this section, we describe our framework. As shown in Figure~\ref{fig:workflow}, the goal of distributed acoustic scene understanding is to convert raw multi-channel acoustic signals and geometric information into a textual description of the scene.

\subsection{Input Formalization via Geometry-Informed Graph}

We define the input to our framework as $\mathcal{I}=\{\bm{\mathcal{X}}, \mathbf{P}, \Omega \}$, which describes the indoor acoustic environment. Specifically, the acoustic data $\bm{\mathcal{X}} \in \mathbb{R}^{N \times T \times F}$ consists of synchronized log-Mel spectrograms from $N$ microphone nodes. Each spectrogram contains $T$ time frames and $F$ frequency bins. The set $\mathbf{P} = \{ \mathbf{p}_1, \dots, \mathbf{p}_N \}$ means the three-dimensional coordinates $\mathbf{p}_n = (x_n, y_n, z_n)$ for each microphone. The environmental geometry $\Omega$ describes the environmental layout and its acoustic properties. We express it as $\Omega = \{ \mathcal{B}, \mathcal{D}, \mathcal{M} \}$. $\mathcal{B}$ gives the room boundaries and the room size. $\mathcal{D}$ records the positions and states of connections between rooms. $\mathcal{M}$ contains material properties such as the absorption coefficient.

For simpler deployments, $\Omega$ may be reduced to room membership and adjacency. Coordinates and attenuation are used here for controlled simulation. Based on the above definition, we represent the multi-room environment as a topological graph $\mathcal{G}_{topo} = (\mathcal{V}, \mathcal{E})$. Each vertex $v \in \mathcal{V}$ corresponds to a physical microphone node. Edges in $\mathcal{E}$ are created between nodes located in the same room or in adjacent rooms connected by a portal. The edge weights further encode spatial distance and transmission properties. We calculate edge weights $w_{ij}$ to model the acoustic propagation between node $i$ and node $j$ as:
\begin{equation}
    w_{ij} = \gamma_{wall} \cdot \exp(-\lambda d_{ij}),
\end{equation}
where $d_{ij} = || \mathbf{p}_i - \mathbf{p}_j ||_2$ is the Euclidean distance and $\lambda$ is a distance-dependent decay constant. The value of the attenuation coefficient $\gamma_{wall}$ is determined by the physical barriers between nodes as specified in $\Omega$. Specifically, $\gamma_{wall} = 1$ if nodes $i$ and $j$ are located in the same room. If they are in adjacent rooms connected by a portal, $\gamma_{wall}$ is determined by the portal state and transmission properties. If separated by walls without a portal, $\gamma_{wall}$ is assigned a value in $(0,1)$ according to the material absorption properties $\mathcal{M}$. This weighting scheme allows the graph to represent non-line-of-sight characteristics and signal leakage across room boundaries.

\subsection{Spatial-Temporal Graph for Distributed Fusion}

Given the acoustic topology and input definitions, the following fusion module synthesizes the synchronized audio streams $\bm{\mathcal{X}}$ with $\mathcal{G}_{topo}$ to extract a unified spatio-temporal representation. The architecture employs a hierarchical pipeline, progressing from localized node-level encoding to relational graph-based message passing.

\textbf{Acoustic Feature Extraction.} Firstly, each microphone node $n$ processes a fixed-length log-Mel segment centered at time index $t$, denoted as $\mathbf{x}_{n,t} \in \mathbb{R}^{\tau \times F}$. Here, $\tau$ is the local segment length, and $t$ denotes the segment index after temporal segmentation of the full spectrogram. We employ the Audio Spectrogram Transformer (AST)~\cite{gong21b_interspeech} as the backbone for our shared encoder $f_{\theta}$. The latent embedding $\mathbf{h}_{n,t} \in \mathbb{R}^{D}$ is extracted from the final block output (i.e., the hidden states before the final classification head). Because the encoder weights $\theta$ are the same across all nodes, the model learns a consistent representation of acoustic events across different rooms. We then concatenate each embedding $\mathbf{h}_{n,t}$ with an MLP-based positional encoding derived from the node coordinates $\mathbf{p}_n \in \mathbf{P}$. This step integrates spatial information into the acoustic features to obtain the initial node states in $\mathcal{G}_{topo}$.

\textbf{Spatial Message Passing.} To model the acoustic coupling between rooms, we first apply a spatial graph neural network at each time step. The graph takes the latent embeddings $\mathbf{H}_t = \{ \mathbf{h}_{1,t}, \dots, \mathbf{h}_{N,t} \}$ as node inputs and the weights $w_{ij}$ from $\mathcal{G}_{topo}$ as edge attributes. For each node $n$, the hidden state $\mathbf{z}_{n,t}$ is updated by aggregating information from its neighbors $j \in \mathcal{N}(n)$ through a graph convolution operation:
\begin{equation}
    \mathbf{z}_{n,t} = \sigma \left( \mathbf{W}_{s} \mathbf{h}_{n,t} + \sum_{j \in \mathcal{N}(n)} w_{nj} \mathbf{W}_{n} \mathbf{h}_{j,t} \right),
\end{equation}
where $\mathbf{W}_{s}$ and $\mathbf{W}_{n}$ are learnable weight matrices for self-representation and neighbor aggregation. The edge weight $w_{nj}$ explicitly scales the contribution of neighboring nodes based on the distance and wall properties defined in $\Omega$. This allows the network to prioritize signals from nodes with direct paths.

\textbf{Temporal Dynamics Modeling.} The spatially-fused vectors $\mathbf{z}_{n,t}$ are then passed through a Gated Recurrent Unit (GRU)~\cite{chung2014empirical} to capture temporal dependencies across time. This stage captures the continuity of sound events and person movements across the multi-room environment. The output of this module is a sequence of spatio-temporal embeddings $\mathbf{Z} = \{ \mathbf{z}_{n,t} \}_{n=1, t=1}^{N, T}$, which serves as the input for the subsequent semantic decoding stage.

\subsection{Semantic Tokenization via Dynamic Acoustic Scene Graph}

Following the generation of spatio-temporal embeddings $\mathbf{Z}$, the framework translates these latent features into a discrete symbolic format, i.e., the dynamic acoustic scene graph. It serves as a semantic bottleneck that extracts task-relevant events.

At each time step $t$, the dynamic acoustic scene graph is instantiated as a set of $K$ semantic triplets $\mathcal{T}_t = \{ \langle s, r, o \rangle_k \}_{k=1}^K$. Each triplet consists of a subject $s \in \mathcal{S}$ (e.g., \textit{Speaker}), a spatial-acoustic relation $r \in \mathcal{R}$ (e.g., \textit{entering}), and an object or location $o \in \mathcal{O}$ (e.g., \textit{Meeting Room}). The scene logic decoder $\Psi$ identifies these triplets by treating the node states $\{\mathbf{z}_{n,t}\}$ as input tokens. In detail, we implement $\Psi$ using a query-based multi-head attention decoder that attends to relevant acoustic regions across rooms and predicts the subject, relation, and location components with separate classifiers.

We factorize the probability of a triplet sequence into a conditional chain to maintain semantic consistency. The joint distribution is defined as:
\begin{equation}
P(\mathcal{T}_t \mid \mathbf{Z}_t)
= \prod_{k=1}^K P(s_k \mid \mathbf{Z}_t)\,
  P(r_k \mid s_k, \mathbf{Z}_t) P(o_k \mid s_k, r_k, \mathbf{Z}_t),
\end{equation}
where the prediction of the relation $r_k$ depends on the identified subject $s_k$, and the object $o_k$ depends on both. This hierarchy prevents logically impossible pairings, such as a ``door" performing a ``speech" action. The geometry $\Omega$ further constrains the location set $\mathcal{O}$ to ensure the predicted movements align with the physical room settings. We supervise the tokenization process using the ground-truth logs from our simulator. During training, the model minimizes a cross-entropy loss over the predicted triplet components and the simulator's event stream. The resulting sequence of triplets $\mathcal{Q} = [\mathcal{T}_1, \dots, \mathcal{T}_T]$ provides a structured and symbolic summary of the scene.

\subsection{Narrative Synthesis via LLM}

The final stage converts the symbolic sequence $\hat{\mathcal{Q}}$ into a natural language narrative $\mathcal{Y}$. Instead of fine-tuning the language model, we use a frozen pre-trained Meta Llama-3-8B-Instruct model\footnote{\href{https://huggingface.co/meta-llama/Meta-Llama-3-8B-Instruct}{Meta-Llama-3-8B-Instruct}} to generate the scene description~\cite{llama3modelcard}.

\textbf{Prompt Construction and Linearization.} The input to the language model is a structured prompt that combines the environment and the detected events. First, the geometry $\Omega$ is converted into a textual description of the floor plan (e.g., \texttt{"Kitchen is adjacent to Meeting Room"}). Second, the sequence $\hat{\mathcal{Q}} = [\hat{\mathcal{T}}_1, \dots, \hat{\mathcal{T}}_T]$ is linearized into a chronological log, where each triplet $\langle s,r,o \rangle$ is formatted as \texttt{"[Time $t$] Subject: Relation: Object"}. The prompt also instructs the LLM to respect adjacency constraints when synthesizing a continuous story.

\textbf{Zero-Shot Temporal Reasoning via LLM.} During inference, the LLM acts as a reasoning engine that resolves spatial ambiguities. For instance, if the symbolic log $\hat{\mathcal{Q}}$ indicates a ``vacuuming" sound in the Meeting Room followed by ``footsteps" in the Corridor, the LLM infers a transition even if the specific "walking" event was partially occluded. This capability for gap-filling is expressed by the generation objective:
\begin{equation}
    \mathcal{Y}^* = \arg\max_{\mathcal{Y}} \prod_{i=1}^L P_{LLM}(y_i \mid y_{<i}, \text{Prompt}(\hat{\mathcal{Q}}, \Omega)),
\end{equation}
where the LLM parameters remain fixed. This strategy ensures that the generated narrative $\mathcal{Y}$ is not a mere list of detected sounds, but a logically consistent and linguistically fluent description of the complete indoor activities.

\section{Experiment}
\label{sec:experiments}

\subsection{Experimental Setup}
\label{sec:exp_setup}

\textbf{Simulator and Dataset}: We evaluate our framework using a custom simulator based on \texttt{pyroomacoustics}~\cite{scheibler2018pyroomacoustics}, extended to support JSON-format floor plans and heterogeneous microphone types. This is a controlled simulation-only study. We simulated multi-room layouts (2 to 4 rooms) with reverberation times (RT60) between 0.2 seconds and 0.6 seconds. We sampled the acoustic events mainly from human speech from \texttt{LibriSpeech}~\cite{panayotov2015librispeech} and environmental sounds from \texttt{ESC-50}~\cite{piczak2015dataset}, all resampled to 16 kHz. Audio clips were randomly cropped or repeated to match event durations, with a maximum concurrency $K$ aligned with the number of decoder queries. A network of $N=6$ distributed microphone nodes was deployed across each environment. Background noise levels varied between 20 and 40 dB SNR, and event trajectories were generated using room-constrained random waypoint sampling to test spatial tracking and reasoning.

\textbf{Evaluation Metrics}: (1) Triplet F1-score ($F_1$) assesses acoustic event detection, requiring a match of the $\langle s,r,o \rangle$ tuple per one second. (2) Standard natural language generation metrics include BLEU-4~\cite{papineni2002bleu}, ROUGE-L~\cite{lin2004rouge}, and BERTScore~\cite{zhang2019bertscore}. (3) Spatial Consistency Score (SCS). For consecutive locations $\hat{o}_t$ and $\hat{o}_{t+1}$, a transition is valid if the locations are identical or adjacent in $\Omega$. SCS measures final narrative consistency and complements Triplet $F_1$.

\subsection{Comparative Evaluation of System Variants}
We conduct a comparative analysis across different sensing paradigms and architectural ablations to evaluate the system's robustness against non-line-of-sight occlusions and its capability for geometry-informed description.

The results shown in Table~\ref{tab:main_results} reveal a substantial performance gap between centralized and distributed sensing paradigms. The Centralized-Single baseline, restricted by its reliance on a single microphone array, achieves a Triplet $F_1$-score of only 0.51, primarily due to its inability to capture acoustic evidence in non-line-of-sight regions. In contrast, our proposed framework achieves a 0.87 Triplet $F_1$-score, demonstrating that distributed spatial diversity is essential for comprehensive multi-room scene understanding. Even without geometry-aware graph construction and geometry-conditioned prompting, the Distributed Acoustic-Only baseline improves the detection accuracy to 0.74, yet it still struggles with spatial ambiguities. By integrating the topology-aware graph and geometry-informed LLM generation, the full framework effectively leverages spatial redundancy to overcome occlusions. This ensures high detection accuracy across complex layouts where single-point sensing typically fails to maintain signal continuity.

\begin{table}[t]
\centering
\caption{Main Performance Comparison across Different System Variants. All metrics should be as high as possible.}
\label{tab:main_results}
\renewcommand{\arraystretch}{1.2}
\resizebox{\columnwidth}{!}{
\begin{tabular}{lccccc}
\toprule
\textbf{System variant} & \textbf{Triplet $F_1$} & \textbf{BLEU-4} & \textbf{ROUGE-L} & \textbf{BERT} & \textbf{SCS (\%)} \\ \midrule
Centralized-Single & 0.51 & 0.15 & 0.22 & 0.45 & 35.0 \\
Distributed Acoustic-Only  & 0.74 & 0.41 & 0.47 & 0.65 & 61.8 \\
Template-Based     & 0.86 & 0.39 & 0.46 & 0.61 & 79.5 \\ \midrule
\rowcolor{gray!16} \textbf{Ours} & \textbf{0.87} & \textbf{0.55} & \textbf{0.62} & \textbf{0.77} & \textbf{88.2} \\ \bottomrule
\end{tabular}}
\end{table}

\subsection{Geometry-Informed Reasoning Analysis}
A key observation is the synergy between the symbolic scene graph and LLM-driven reasoning. The proposed framework yields a significant improvement in linguistic metrics, achieving a BLEU-4 of 0.55 and a ROUGE-L of 0.62, which far exceeds the Template-Based approach. This suggests that the final narrative benefits from both acoustic triplet evidence and topology-aware LLM priors. Triplet $F_1$ and SCS should therefore be interpreted together. In scenarios where acoustic signals are intermittent or muffled, the LLM leverages room connectivity in the prompt to bridge fragmented detections. Furthermore, the SCS of 88.2\% confirms that geometry information reduces physically impossible transitions in the generated narratives. This grounding mechanism helps the narrative remain both linguistically fluent and physically plausible within the simulated setting.

\subsection{Ablation Study on Architectural Components}

The contribution of each architectural component is further discussed in the ablation study presented in Table~\ref{tab:ablation}. Different from the Distributed Acoustic-Only baseline in Table~\ref{tab:main_results}, the w/o Geometry Prior variant keeps the proposed framework but removes only the explicit geometric prior $\Omega$. It leads to the most drastic decline in spatial logic, with the SCS falling to 66.5\%, showing that acoustic evidence without explicit layout constraints is insufficient for reliable multi-room scene description. Besides, without spatial graph-based message passing, each microphone works in isolation. This prevents the model from combining sound cues from different locations, leading to fragmented graphs and reducing the Triplet $F_1$-score to 0.81. These results show that spatial-temporal graph fusion is essential for building a consistent global representation from scattered data.

\begin{table}[t]
\centering
\caption{Ablation Study on Architectural Components. All metrics should be as high as possible.}
\label{tab:ablation}
\renewcommand{\arraystretch}{1.2}
\resizebox{\columnwidth}{!}{
\begin{tabular}{lccc}
\toprule
\textbf{Configuration} & \textbf{Triplet $F_1$} & \textbf{BERT} & \textbf{SCS (\%)} \\ \midrule
\rowcolor{gray!15} \textbf{Our Full Framework} & \textbf{0.87} & \textbf{0.77} & \textbf{88.2} \\
\quad -- w/o Geometry Prior ($\Omega$) & 0.78 & 0.68 & 66.5 \\
\quad -- w/o ST-Graph Fusion & 0.81 & 0.70 & 80.9 \\ \bottomrule
\end{tabular}}
\vspace{-4pt}
\end{table}

\section{Conclusion}

In this paper, we introduced a distributed framework for understanding acoustic scenes in multi-room settings. We mitigate sound occlusion using distributed microphones and graph-based fusion, and use geometry-informed LLM reasoning to reduce non-physical transitions in the final narrative. The results demonstrate the promise of distributed sensing for complex indoor environments under controlled simulation. Future work will validate the framework in real-world physical testbeds and optimize inference speed.

\newpage
\clearpage
\section{Use of Generative AI Disclosure}
The authors used a generative AI tool solely for language editing and polishing purposes. The AI tool did not contribute to the scientific content, experimental design, data analysis, or conclusions of this manuscript. All authors take full responsibility for the content of the paper.

\bibliographystyle{IEEEtran}
\bibliography{6_reference}

\newpage
\appendix
\clearpage
\section{Limitations and Future Work}
\label{sec:appendix_limitations}

Despite the promising simulation results, three limitations remain. (1) \textbf{Sim-to-Real Gap}: The evaluation uses known floor plans, microphone positions, and door states. Physical buildings additionally introduce device mismatch, imperfect synchronization, changing furniture, and non-stationary background noise. Future work will validate the framework in multi-room testbeds. (2) \textbf{Scene Complexity and Uncertainty}: The generated data are relatively structured and do not yet cover dense, overlapping sources in the same room (the ``cocktail party'' effect). Moreover, incorrect or uncertain triplets may lead the LLM to produce a confident but incorrect narrative. Future systems should propagate prediction uncertainty and distinguish directly detected events from inferred transitions. (3) \textbf{Static Topology}: The framework assumes a known, static $\Omega$. Dynamic doors or temporary partitions require adaptive updates to topology and propagation priors.

\section{Details of the Scene Logic Decoder Architecture} 

The scene logic decoder $\Psi$ is designed to translate the high-dimensional graph states $\mathbf{Z}_t$ into a set of discrete semantic triplets. This architecture utilizes a query-based Transformer mechanism to detect multiple concurrent acoustic events within a multi-room environment.

\textbf{Query-based Detection Mechanism.} We define a set of $K$ learnable query embeddings $\mathbf{U} = \{\mathbf{u}_1, \dots, \mathbf{u}_K\}$, where $K$ is the maximum number of simultaneous events allowed per time step. Each query $\mathbf{u}_k$ acts as a semantic probe that searches for specific acoustic signatures across the graph nodes. Through a multi-head cross-attention mechanism, the queries interact with the spatio-temporal embeddings $\mathbf{Z}_t = \{\mathbf{z}_{1,t}, \dots, \mathbf{z}_{N,t}\}$:
\begin{equation}
\begin{split}
\mathbf{f}_k 
&= \text{MultiHeadAttn}(\mathbf{u}_k, \mathbf{Z}_t, \mathbf{Z}_t) \\
&= \text{Softmax}\left(
\frac{(\mathbf{u}_k \mathbf{W}_Q)
(\mathbf{Z}_t \mathbf{W}_K)^\top}
{\sqrt{d_{\mathrm{att}}}}
\right)
(\mathbf{Z}_t \mathbf{W}_V),
\end{split}
\end{equation}
where $\mathbf{W}_Q$, $\mathbf{W}_K$, and $\mathbf{W}_V$ are learnable projections and $d_{\mathrm{att}}$ is the attention scaling dimension. The resulting vector $\mathbf{f}_k$ represents an event-level acoustic feature, integrating the ``what'' from the AST encoder and the ``where'' from the spatial graph.

\textbf{Hierarchical Triplet Classification.} To produce the final symbolic tokens, $\mathbf{f}_k$ is passed through three prediction heads. The subject is predicted first, the relation is conditioned on the subject, and the object or room is conditioned on both the subject and relation:
\begin{equation}
    \hat{s}_k = \text{Softmax}(\mathbf{W}_s \mathbf{f}_k),
\end{equation}
\begin{equation}
    \hat{r}_k = \text{Softmax}(\mathbf{W}_r [\mathbf{f}_k; \text{Emb}(s_k^{gt})]),
\end{equation}
\begin{equation}
    \hat{o}_k = \text{Softmax}(\mathbf{W}_o [\mathbf{f}_k; \text{Emb}(s_k^{gt}); \text{Emb}(r_k^{gt})]),
\end{equation}
where $\text{Emb}(\cdot)$ denotes a small label-embedding layer. This structure prevents invalid combinations such as a ``Door'' performing ``Speech''. If fewer than $K$ events occur, the remaining queries predict a special \texttt{[NULL]} token. During training, teacher forcing conditions the relation and object heads on the ground-truth subject and relation; during inference, predicted labels are used autoregressively.

\textbf{Training Objective.} The decoder is supervised using a cross-entropy loss applied to each triplet component. During training, predicted queries are matched with ground-truth events using a fixed ordering based on event timestamps, ensuring consistent supervision. Given the ground-truth triplets $\mathcal{T}_t^{gt}$ provided by the simulator, the loss function is defined as:
\begin{equation}
    \mathcal{L} = \sum_{t=1}^T \sum_{k=1}^K \left( \mathcal{L}_{CE}(s_k, \hat{s}_k) + \mathcal{L}_{CE}(r_k, \hat{r}_k) + \mathcal{L}_{CE}(o_k, \hat{o}_k) \right).
\end{equation}
By minimizing this objective, the decoder learns to map specific spatial-acoustic patterns (e.g., high-frequency transients in the kitchen) to their corresponding semantic labels (e.g., \textit{Subject: Knife, Relation: Chopping, Object: Kitchen}). This transformation effectively compresses the raw audio evidence into a structured format for the subsequent LLM-driven reasoning and generation.

\section{LLM Prompt Construction}
\label{sec:appendix_prompt}

The inference process $\mathcal{Y}^* = \arg\max_{\mathcal{Y}} \prod_{i=1}^L P_{LLM}(y_i \mid y_{<i}, \text{Prompt}(\hat{\mathcal{Q}}, \Omega))$ relies on two inference-time inputs: environmental geometry $\Omega$ and predicted acoustic observations $\hat{\mathcal{Q}}$. Simulator event logs and reference narratives are excluded from the prompt and used only for training supervision and evaluation, respectively. Table~\ref{tab:prompt_multi_agent} shows their linearization.

\begin{table}[!ht]
\centering
\caption{Example prompt for geometry-informed reasoning. The LLM receives $\Omega$ and predicted triplets $\hat{\mathcal{Q}}$; reference data are excluded.}
\label{tab:prompt_multi_agent}
\small
\renewcommand{\arraystretch}{1.5}
\begin{tabularx}{\columnwidth}{|X|}
\hline
\rowcolor{gray!15} \textbf{[System Instruction]} \\
You are an expert in indoor spatial reasoning. Your task is to synthesize fragmented acoustic triplets from multiple rooms into a physically plausible and continuous narrative. \\ \hline
\rowcolor{gray!15} \textbf{[Environmental Layout $\Omega$]} \\
The building layout is defined by the following room adjacencies: \\
- \{\textbf{Living Room}\} is connected to: \{\textbf{Corridor}\}. \\
- \{\textbf{Corridor}\} is connected to: \{\textbf{Living Room, Kitchen}\}. \\
- \{\textbf{Kitchen}\} is connected to: \{\textbf{Corridor}\} (via a door). \\
\textit{Constraint:} A person or an agent cannot move between non-adjacent rooms without passing through the connecting space.  \\ \hline
\rowcolor{gray!15} \textbf{[Predicted Acoustic Observations $\hat{\mathcal{Q}}$]} \\
The distributed sensing network detected the following concurrent events: \\
1. [0--3s] $\langle$Person, footstep, Living Room$\rangle$; $\langle$Person, preparing food, Kitchen$\rangle$ \\
2. [3--6s] $\langle$Person, footstep, Corridor$\rangle$; $\langle$Person, preparing food, Kitchen$\rangle$ \\
3. [6--7s] $\langle$Person, opening door, Corridor$\rangle$; $\langle$Person, preparing food, Kitchen$\rangle$ \\
4. [7--10s] $\langle$Person, talking, Kitchen$\rangle$; $\langle$Person, preparing food, Kitchen$\rangle$ \\ \hline
\rowcolor{gray!15} \textbf{[Reasoning Task]} \\
Synthesize a continuous story. If a spatial gap exists in $\hat{\mathcal{Q}}$, use $\Omega$ to infer the most likely missing transitions. Ensure that the output follows the floor plan. \\ \hline
\end{tabularx}
\end{table}

\section{Extended Experimental Configurations}
\label{sec:appendix_exp}

\subsection{Acoustic Simulator and Environment Modeling}

The simulation platform is built upon the \texttt{pyroomacoustics}\footnote{\url{https://github.com/LCAV/pyroomacoustics}} library and adapted for multi-room distributed acoustic sensing scenarios.

\begin{itemize}
    \item \textbf{JSON-based Topology}: Floor plans are defined via structured JSON files specifying room size and shape, portal (door) connections, and material properties. This configuration enables flexible generation of multi-room environments while preserving explicit adjacency relationships used in spatial reasoning.

    \item \textbf{Dynamic Sound Sources}: Sound sources (i.e., speech or environmental events) are allowed to move over time. For each event, the simulator samples spatial trajectories constrained by room boundaries.

    \item \textbf{Acoustic Propagation}: Acoustic propagation is simulated using the Image Source Method (ISM) provided by \texttt{pyroomacoustics}. Frequency-dependent wall absorption coefficients are specified according to material types. Reverberation characteristics are controlled by sampling RT60 values within the 0.2--0.6 s range.

    \item \textbf{Heterogeneous Microphones}: Each environment can deploy multiple and multi-type microphone nodes. The microphones are spatially distributed across different rooms to emulate distributed sensing conditions.

    \item \textbf{Structured Ground Truth Output}: The simulator automatically exports event annotations in structured JSON format, including subject, spatial relation, and room location.
\end{itemize}

For the three-room example, $\mathcal{B}$ contains a $4.0\times4.0\times3.0$~m living room, a $4.0\times2.0\times3.0$~m corridor, and a $4.0\times4.0\times3.0$~m kitchen. $\mathcal{D}$ records an open living-room--corridor connection at $(4.0,2.0,0.0)$ and an initially closed kitchen door at $(8.0,2.0,0.0)$. $\mathcal{M}$ assigns absorption coefficients of approximately 0.21, 0.25, and 0.18, corresponding to target RT60 values of 0.45, 0.30, and 0.55~s, respectively.

\subsection{Dataset Synthesis and Pre-processing}
To construct the training and test sets, we performed a multi-stage synthesis:
\begin{itemize}
    \item \textbf{Polyphase Resampling}: Since \texttt{ESC-50} is recorded at 44.1 kHz, we apply a filter to downsample all environmental clips to 16 kHz. This ensures spectral alignment with the \texttt{LibriSpeech} dataset.
    \item \textbf{Spatio-Temporal Mixing}: For each scene, we sample $K$ active events. Each event is convolved with a time-varying room impulse response corresponding to its trajectory relative to the $N$ microphone nodes.
\end{itemize}

The simulator stores an event log and a reference narrative for each evaluation scene. For the three-room example in Table~\ref{tab:prompt_multi_agent}, the event log contains Person A walking in the living room from 0 to 3~s, Person B preparing food in the kitchen from 0 to 10~s, Person A walking through the corridor from 3 to 6~s, Person A opening the kitchen door from 6 to 7~s, and both people talking in the kitchen from 7 to 10~s. The corresponding reference is:
\begin{quote}
\textit{``Person B is preparing food in the kitchen for the whole ten-second recording. During the first three seconds, Person A walks around the living room. Person A then walks along the corridor between about 3 and 6 seconds, and opens the door at the kitchen end of the corridor at 6 seconds. After entering the kitchen at about 7 seconds, Person A talks with Person B in the kitchen until the end of the recording.''}
\end{quote}
The event log supplies triplet-level supervision, while the reference narrative is held out from the LLM prompt and used only by the language-generation metrics.

\subsection{Formal Definitions of Evaluation Metrics}

\textbf{Triplet F1-score ($F_1$)}: To assess the system's ability to capture structured semantic information, we employ a \textit{segment-based F1-score} with a temporal resolution of 1 second. For each 1-second segment $t$, a predicted triplet $\hat{\mathcal{T}}_t = \langle \hat{s}, \hat{r}, \hat{o} \rangle$ is considered a True Positive ($TP$) if and only if all three components, (i.e., subject, relation, and object), match the ground truth exactly, i.e., $(\hat{s}, \hat{r}, \hat{o}) = (s_{gt}, r_{gt}, o_{gt})$. Precision ($P$) and Recall ($R$) are calculated by aggregating $TP$, False Positives ($FP$), and False Negatives ($FN$) across all segments. The micro-averaged $F_1$ score is defined as:
\begin{equation}
    F_1 = \frac{2 \cdot P \cdot R}{P + R},
\end{equation}
where $P = \frac{\sum TP}{\sum TP + \sum FP}$ and $R = \frac{\sum TP}{\sum TP + \sum FN}$.

\textbf{Standard Natural Language Generation (NLG) Metrics}:
To evaluate the linguistic quality and semantic faithfulness of the generated narrative $\mathcal{Y}$ against the ground-truth narrative $\mathcal{Y}_{gt}$, we employ the following metrics:

\begin{itemize}
    \item \textbf{BLEU-4}: Measures $n$-gram precision ($n=1 \dots 4$) with a brevity penalty ($BP$) to penalize short generations. It is defined as:
    \begin{equation}
        \text{BLEU} = BP \cdot \exp \left( \sum_{n=1}^4 w_n \log p_n \right),
    \end{equation}
    where $p_n$ is the precision of $n$-grams and $w_n$ are uniform weights.
    
    \item \textbf{ROUGE-L}: Evaluates the Longest Common Subsequence (LCS) between $\mathcal{Y}$ and $\mathcal{Y}_{gt}$. It accounts for sentence-level structure by identifying the longest sequence of words that appear in both texts in the same relative order. The $F$-measure is calculated based on LCS-based recall ($R_{lcs}$) and precision ($P_{lcs}$):
    \begin{equation}
        F_{lcs} = \frac{(1 + \beta^2) R_{lcs} P_{lcs}}{R_{lcs} + \beta^2 P_{lcs}},
    \end{equation}
    where we set $\beta = 1$ following standard ROUGE-L settings.
    \item \textbf{BERTScore}: Unlike $n$-gram based metrics, BERTScore captures semantic similarity using contextual embeddings from a pre-trained BERT model. We compute the cosine similarity between the embedding vectors of each token in $\mathcal{Y}$ and $\mathcal{Y}_{gt}$, followed by greedy matching to maximize the total similarity score:
    \begin{equation}
        R_{BERT} = \frac{1}{|\mathcal{Y}_{gt}|} \sum_{x_i \in \mathcal{Y}_{gt}} \max_{\hat{x}_j \in \mathcal{Y}} \mathbf{x}_i^\top \hat{\mathbf{x}}_j,
    \end{equation}
    where $\mathbf{x}_i$ and $\hat{\mathbf{x}}_j$ are the pre-normalized contextual embeddings of tokens in the reference and candidate narratives, respectively.
\end{itemize}

\textbf{Spatial Consistency Score (SCS)}:
The SCS quantifies the physical plausibility of the generated narrative. Let $\tilde{\mathbf{A}}\in\{0,1\}^{N_{\mathrm{room}}\times N_{\mathrm{room}}}$ be the adjacency-with-self matrix derived from $\Omega$, where $\tilde{A}_{ij}=1$ if $i=j$ or if rooms $i$ and $j$ are connected by a valid portal. For each subject-specific location sequence $O^s=[\hat{o}^{\,s}_1,\ldots,\hat{o}^{\,s}_{L_s}]$, let $\mathcal{S}_{\mathrm{eval}}$ contain all sequences with $L_s\geq2$ across the evaluated scenes. The transition-level micro-average is
\begin{equation}
\mathrm{SCS}=100\times
\frac{\displaystyle\sum_{s\in\mathcal{S}_{\mathrm{eval}}}\sum_{t=1}^{L_s-1}
\mathbb{I}(\tilde{A}_{\hat{o}^s_t,\,\hat{o}^s_{t+1}}=1)}
{\displaystyle\sum_{s\in\mathcal{S}_{\mathrm{eval}}}(L_s-1)},
\end{equation}
where $\mathbb{I}(\cdot)$ is the indicator function. Sequences with fewer than two locations have no transition and are excluded; if none remain, SCS is undefined and is not reported.

\subsection{Hyperparameters and Deployment}

\textbf{Feature Extraction}:
The raw audio is processed at a 16 kHz sampling rate. For the Short-Time Fourier Transform (STFT), we use a Hann window with a hop length of 128 samples. Spectrograms are standardized and trimmed to a fixed dimension of $T = 256$ frames.

\textbf{Implementation and Hardware}:
The framework is implemented in PyTorch under Python 3.8. Training is conducted on NVIDIA A10 (24GB) GPU. We use the Adam optimizer with an initial learning rate of 0.0005, decaying to 95\% every 10 epochs. We employ a batch size of 32 and train the pipeline for 80 epochs.

\textbf{Scaling to Longer Recordings and Larger Venues}:
The framework can operate in overlapping temporal windows of $T=256$ frames, merging triplets by timestamp and subject identity. Large floor plans can be decomposed into acoustically connected subgraphs with overlapping boundary nodes around doors and corridors; local triplet logs are then reconciled into a building-level chronology. Practical limits are set by the microphones and edges per graph window, temporal context, and the LLM context needed to serialize the relevant geometry. The controlled study uses 2--4 rooms and $N=6$ microphones.

\subsection{Experimental Objectives and Baseline Design}
\label{sec:appendix_objectives}
The experimental evaluations in the main body Table 1 and Table 2 are designed to validate the system across three dimensions: sensing paradigm, fusion efficiency, and reasoning logic.

Main Performance Comparison (baselines). (1) \textbf{Centralized vs. Distributed}: We compare a single-room microphone array (\textit{Centralized-Single}) with \textit{Distributed Acoustic-Only}, which retains distributed microphones but removes geometry-aware graph construction and geometry-conditioned prompting. This tests the value of spatial diversity under non-line-of-sight occlusion. (2) \textbf{Topology-Aware Fusion}: The comparison between \textit{Distributed Acoustic-Only} and the full framework tests whether explicit propagation and topology improve the use of distributed observations. (3) \textbf{Deterministic vs. Reasoning-based Output}: \textit{Template-Based} maps detected triplets to text without LLM reasoning, separating verbalization from geometry-informed gap filling. Together, the variants isolate distributed sensing, topology-aware fusion, and narrative reasoning.

Ablation Study. (1) \textbf{Impact of Geometry Prior ($\Omega$)}: Unlike \textit{Distributed Acoustic-Only}, \textit{w/o Geometry Prior} keeps the proposed pipeline and removes only the explicit $\Omega$, isolating its contribution to spatial plausibility. (2) \textbf{Role of Graph Fusion}: \textit{w/o ST-Graph Fusion} removes graph-based message passing to test spatio-temporal aggregation while retaining the rest of the framework.

\section{Structured Data Format}
\label{sec:appendix_json}

Each simulated scene is stored as a structured JSON object containing room geometry $\Omega$, microphone coordinates, portal states, material parameters, and simulator event metadata. The file supports simulation and supervision; at inference, the LLM receives only serialized $\Omega$ and predicted triplets $\hat{\mathcal{Q}}$, not the event log or reference narrative. A representative scene file is shown below:

\begin{lstlisting}[language=json, caption={A simplified JSON scene with six microphones and physical constraints.}]
{
  "building": {
    "rooms": [
      {
        "room_id": "living_room",
        "dimension": [4.0, 4.0, 3.0],
        "rt60": 0.45,
        "absorption": 0.21,
        "mics": [
          {"id": "mic_1", "pos": [1.0, 2.0, 1.5]},
          {"id": "mic_2", "pos": [3.0, 2.0, 1.5]}
        ]
      },
      {
        "room_id": "corridor",
        "dimension": [4.0, 2.0, 3.0],
        "rt60": 0.30,
        "absorption": 0.25,
        "mics": [
          {"id": "mic_3", "pos": [5.0, 1.0, 1.2]},
          {"id": "mic_4", "pos": [7.0, 1.0, 1.2]}
        ]
      },
      {
        "room_id": "kitchen",
        "dimension": [4.0, 4.0, 3.0],
        "rt60": 0.55,
        "absorption": 0.18,
        "mics": [
          {"id": "mic_5", "pos": [9.0, 2.0, 1.5]},
          {"id": "mic_6", "pos": [11.0, 2.0, 1.5]}
        ]
      }
    ]
  },
  "connections": [
    {
      "from": "living_room",
      "to": "corridor",
      "door_position": [4.0, 2.0, 0.0],
      "status": "open"
    },
    {
      "from": "corridor",
      "to": "kitchen",
      "door_position": [8.0, 2.0, 0.0],
      "status": "closed_initially"
    }
  ],
  "events": [
    {
      "type": "footstep",
      "subject": "Person_A",
      "room": "living_room",
      "position": [2.0, 2.0, 0.0],
      "start": 0.0, "end": 3.0,
      "audio_path": "XX"
    },
    {
      "type": "preparing_food",
      "subject": "Person_B",
      "room": "kitchen",
      "position": [10.0, 2.0, 1.0],
      "start": 0.0, "end": 10.0,
      "audio_path": "XX"
    },
    {
      "type": "footstep",
      "subject": "Person_A",
      "room": "corridor",
      "position": [6.0, 1.0, 0.0],
      "start": 3.0, "end": 6.0,
      "audio_path": "XX"
    },
    {
      "type": "door_open",
      "subject": "Person_A",
      "room": "corridor",
      "position": [8.0, 2.0, 1.2],
      "start": 6.0, "end": 7.0,
      "audio_path": "XX"
    },
    {
      "type": "speech",
      "subject": "Person_A_B",
      "room": "kitchen",
      "position": [9.5, 2.0, 1.6],
      "start": 7.0, "end": 10.0,
      "audio_path": "XX"
    }
  ]
}
\end{lstlisting}

\newpage
\section{Qualitative Narrative Comparison}
The following ten-second example compares a reference narrative with outputs from the full framework and the Centralized-Single baseline.

\textbf{Ground truth.} \textit{``At the start of the recording, Person B is talking in the living room, and a dog is barking in the kitchen from 1 to 3 seconds. At around 3 seconds, Person A makes footstep sounds in the bedroom, and then moves into the corridor at around 5 seconds. At 7 seconds, Person A opens the door between the corridor and the living room, enters the living room, and from 8 to 10 seconds Person A and Person B are talking together in the living room.''}

\textbf{Full framework.} \textit{``At around 0 seconds, Person B is talking in the living room, while a dog is barking in the kitchen. At 3 seconds, Person A enters the bedroom, making footstep sounds. They then move into the corridor, still making footstep sounds, at around 5 seconds. At 7 seconds, Person A opens the door between the corridor and the living room. Shortly after, at around 8 seconds, Person A and Person B both enter the living room, where they start talking together.''}

\textbf{Centralized-Single.} \textit{``At around 0 seconds, Person A is talking in the living room. At 1 second, a dog in the bedroom starts barking. At 3 seconds, the barking stops. At 5 seconds, Person A enters the kitchen, making footstep sounds. At 8 seconds, Person A returns to the living room, where they start talking again.''}

The full framework preserves the main event identities, locations, and temporal order, although it omits the barking interval and incorrectly says that both people enter the living room. Centralized-Single changes the initial speaker, puts the dog in the wrong room, redirects Person A's movement, and omits the final joint conversation.

\end{document}